\documentclass{article}
\usepackage{isita2004}
\usepackage{graphicx}
\usepackage{amsmath}

\title{Maximum Mutual Information of Space-Time Block Codes\\with Symbolwise Decodability}

\name{Kenji Tanaka$^{\dagger}$, Ryutaroh Matsumoto$^{\dagger}$ and Tomohiko Uyematsu$^{\dagger}$}

\address{
\begin{tabular}{c}
$^{\dagger}$
Department of Communications and Integrated Systems\\
Tokyo Institute of Technology\\
2-12-1 Ookayama, Meguro-ku, Tokyo, 152-8550, JAPAN\\
E-mail:\{kenji,ryutaroh,uematsu\}@it.ss.titech.ac.jp
\end{tabular}
}

\begin{document}

\maketitle
\sloppy
\begin{abstract}
In this paper, we analyze the performance of space-time block codes which enable symbolwise maximum likelihood
decoding. We derive an upper bound of maximum mutual information (MMI) on
space-time block codes that enable symbolwise maximum likelihood decoding for a frequency non-selective
quasi-static fading channel. MMI is an
upper bound on how much one can send information with vanishing error
probability by using the target code.
\end{abstract}

\def\tr{\mathop{\rm tr}\nolimits}
\def\diag{\mathop{\rm diag}\nolimits}
\newcommand{\argmax}{\mathop{\rm argmax}\limits}

\section{Introduction}\label{intro}
\sloppy
An important problem in future telecommunication systems will be how to
send large amount of data such as video through a wireless channel at
high rate with high reliability in a mobile environment. One way to
enable the high rate communication on the scattering-rich wireless
channel is use of multiple transmit and receive antennas. It is well
known that the capacity of a wireless channel linearly increases as the
number of transmit and receive antennas under the condition that total
power and bandwidth of signals are constant~\cite{Telatar,Foschini}. A wireless
communication system with multiple transmit and receive antennas is
called multi-input multi-output (MIMO) system and an encoding/modulation
method for MIMO system is called space-time code.

In the design of space-time codes, it is desirable to reduce the size
of the circuit for encoding and decoding. Jiang et al.~\cite{Jiang} and
Khan et al.~\cite{Khan} derived a necessary and sufficient condition for
symbolwise maximum likelihood (ML) decoding  on linear dispersion codes (LDC)~\cite{Hassibi}, and such code
design is named single symbol decodable design (SSDD). Complex linear
processing orthogonal design (CLPOD)~\cite{Tarokh}, which is the
subclass of SSDD, has full diversity and maximum
coding gain because of an additional condition on codewords. Many
researchers have studied concrete construction~\cite{Liang},
rate~\cite{Wang}, BER  and the capacity utilization efficiency~\cite{Hassibi,Sandhu} of
CLPOD.

The capacity utilization efficiency of a code can be measured by calculating
its attainable maximum mutual information (MMI). 
MMI of a code is defined as the capacity of a channel which consists of
an encoder of the target code and the original channel, so it is an
upper bound on how much one can send information with vanishing error
probability by using the code. Therefore we know how much a space-time
block code  utilizes the
capacity of wireless channel by calculating their MMI.  MMI is also an important measure of an
inner code of a concatenated code because it corresponds to maximum
possible information rate with vanishing error probability by taking the
block length of an outer code large. Hassibi and Hochwald~\cite{Hassibi} computed MMI of
Alamouti's code~\cite{Alamouti} and MMI of an example of rate-3/4 CLPOD and
mentioned that these
values are far below original channel capacity with more than one receive
antenna. They also proposed LDCs
whose MMI are close to original channel capacity for several numbers of
transmit/receive antennas. Sandhu and Paulraj~\cite{Sandhu} derived the
expression of MMI of general CLPOD and showed that for Rayleigh fading
channel, the value equals to original channel capacity only when one
receive antenna is used. However, there is no knowledge of MMI of SSDD,
which is a subclass of LDC and includes CLPOD as a special case. The
importance of this problem is also mentioned in the
literature~\cite{Jiang}.

In this paper we compute MMI of SSDD over frequency non-selective
quasi-static Rayleigh fading channel and clarify the necessary symbol
rate at which SSDD utilizes full capacity of original channel. This paper is organized as follows: In Section~\ref{Sec.2}, we introduce a mathematical
model for MIMO systems and the definitions of SSDD and CLPOD. In Section~\ref{Sec.3}, we derive an upper bound on MMI of
SSDD. Also we give alternative derivation of the exact expression of MMI
of CLPOD. In Section~\ref{Sec.4}, we show the tightness of the upper
bound on MMI of SSDD by comparing it with delay optimal complex
orthogonal design (COD), which is a subclass of CLPOD at the same symbol
rate. Then we clarify the
necessary symbol rate of SSDD at which MMI of SSDD can attain original channel
capacity. We show that the necessary symbol rate is much
larger than that of CLPOD upper bounded by 3/4. Finally,
Section~\ref{Sec.5} provides our conclusions.

\textit{Notation:}\ Upper case letters denote matrices and bold lower case
letter denote vectors; $\Re(\cdot)$ and $\Im(\cdot)$ denote real and imaginary part of complex number, respectively; $(\cdot)^t$ and $(\cdot)^H$ denote transpose and
Hermitian transpose, respectively; $a_{ij}$ and $x_i$ denote the
$(i,j)$th entry of a matrix $A$ and $i$th entry of a vector ${\bf x}$,
respectively; $I_M$ denotes the identity matrix of size $M$;
$\det(\cdot)$ and $\tr(\cdot)$ denote determinant and trace of a matrix,
respectively; $\diag({\bf x})$ is a diagonal matrix with ${\bf x}$ on its diagonal;
$E_A[\cdot]$ and $E_{\bf x}[\cdot]$ denote expectation over random matrix $A$ and
random vector ${\bf x}$, respectively; covariance matrix of random
vector ${\bf x}$ is denoted as $\Gamma_{\bf x}$. We always index matrix and vector entries
starting from 1. Complex and real field are denoted as ${\bf C}$ and
${\bf R}$, respectively.
\section{Preliminaries}
\label{Sec.2}
In this section, we introduce a mathematical model for MIMO system and define several space-time codes.
\subsection{Mathematical Model for MIMO System}
\sloppy
We consider a communication system that uses $M$ transmit antennas and
$N$ receive antennas. Each transmit antenna simultaneously sends a
narrow band signal through a frequency non-selective Rayleigh fading
channel. The fading is assumed to be quasi-static so that the fading
coefficients are constant for $T$ channel uses. We can write the
relation between a transmitted block (or a codeword) $S$ and a received
block $R$ as follows.
 \begin{eqnarray}
 R&=&\sqrt{\frac{\rho}{M}}SH+V\label{eq:CHANNEL_MODEL1}\\ 
  R&=&[r_{tm}]\in{\bf C}^{T\times N}\nonumber \\
  S&=&[s_{tm}]\in{\bf C}^{T\times M}\nonumber \\
  H&=&[h_{mn}]\in{\bf C}^{M\times N}\nonumber \\
  V&=&[v_{tm}]\in{\bf C}^{T\times N}\nonumber 
 \end{eqnarray}
\begin{center}
 \begin{tabular}{lcl}
  $r_{tn}$ &:& signal to $n$th receive antenna at time $t$\\
  $s_{tm}$ &:& signal from $m$th transmit antenna at time $t$\\
  $h_{mn}$ &:& fading coefficient between $m$th transmit\\
           && antenna and $n$th receive antenna\\
  $v_{tn}$ &:& additive white Gaussian noise (AWGN) at\\
           && $n$th receive antenna at time $t$\\
  $\rho$   &:& SNR at each receive antenna
 \end{tabular}
 \end{center}
\vspace{3mm}
The fading coefficient $h_{mn}$ and AWGN $v_n$ are statistically
independent complex Gaussian random variables with zero mean and unit
variance. We assume that the receiver knows a realization of fading
coefficients (i.e. receiver has perfect CSI (Channel State Information))
but the transmitter does not. The transmitted block is assumed to
satisfy following power constraint:
\begin{equation}
\sum_{t=1}^T\sum_{m=1}^M|s_{tm}|^2=E\left[\tr(SS^*)\right]\leq TM. \label{eq:POWER_CONSTRAINT1}
\end{equation}

\subsection{Definitions of Codes}
\noindent
\textit{Definition(LDC)~\cite{Hassibi}:\ }A linear dispersion code (LDC) is a
space-time code whose codeword $S$ is generated from information sequence
${\bf u}=[u_1,\cdots,u_{2Q}]^t\in {\bf R}^{2Q}$ as
\begin{equation}
S=\sum_{q=1}^{2Q}u_qA_q, \label{eq:LDC}
\end{equation}
where $A_q\in{\bf C}^{T\times M}\ (q=1,\cdots,2Q)$ is called dispersion
matrices.\\

\noindent
\textit{Definition(SSDD)~\cite{Jiang,Khan}:\ }A single symbol decodable
design (SSDD) is an LDC whose dispersion
matrices satisfy the following equations:
\begin{equation}
A_q^HA_r+A_r^HA_q=O_M\ (0\leq q\neq r\leq 2Q). \label{eq:SSDD}
\end{equation}
Jiang et al.~\cite{Jiang} and  Khan et al.~\cite{Khan} proved that a receiver can execute ML decoding for each symbol $u_q$ instead
of each sequence ${\bf u}$ iff Eq.~(\ref{eq:SSDD}) holds. Applying
Eq.~(\ref{eq:LDC}) and Eq.~(\ref{eq:SSDD}) to Eq.~(\ref{eq:POWER_CONSTRAINT1}), we have the
following power constraint on ${\bf u}$ for SSDD:
\begin{equation}
\tr(D_A\Gamma_{\bf u})\leq TM, \label{eq:POWER_CONSTRAINT2} \\
\end{equation}
where $\Gamma_{\bf u}$ denotes the covariance matrix of the random
vector ${\bf u}$ and
\begin{equation}
D_A=\diag\left(\tr(A_1^HA_1),\cdots,\tr(A_{2Q}^HA_{2Q})\right). \label{eq:DIAGONAL1}
\end{equation}
\\

\noindent
\textit{Definition(CLPOD)~\cite{Tarokh}:\ }A complex linear processing orthogonal design
(CLPOD) is an SSDD whose dispersion matrices also satisfy the following
equations:
\begin{equation}
A_q^HA_q=I_M\ (q=1,2,\cdots,2Q). \label{eq:CLPOD}
\end{equation}
It is known that CLPOD achieves full diversity and maximum
coding gain over all LDCs subject to a constant signal constellation and
a constant number of dispersion matrices. For a code in the class of CLPOD, we may use following simple power constraint of on ${\bf u}$:
\begin{equation}
 \tr(\Gamma_{\bf u})\leq T. \label{eq:POWER_CONSTRAINT3}
\end{equation}

\noindent
\textit{Definition(COD)~\cite{Tarokh}:\ }A complex orthogonal design
(COD) is an CLPOD whose dispersion matrices satisfy following
constraint:
\begin{itemize}
\item $A_q\in\{0,\pm 1,\pm j\}^{T\times M}\ (1\leq q\leq 2Q)$
\item If $A_{2q-1}$ or $A_{2q}\ (1\leq q\leq Q)$ has a nonzero $(i,j)$
      entry, then $(i,j)$ entries of $A_{2r-1}$ and $A_{2r}\ (1\leq r\neq q\leq Q)$ are all zero.
\end{itemize}

\section{Maximum Mutual Information (MMI) of SSDD and CLPOD}
\label{Sec.3}
\sloppy
In this section, we discuss capacity utilization of space-time
codes with symbolwise decodability by deriving their MMI. MMI of
CLPOD is derived in~\cite{Sandhu}, but MMI of SSDD is unknown because the key
equation Eq.~(3) in~\cite{Sandhu} does not hold for general SSDD. We
give an upper bound on MMI of SSDD and alternative derivation of exact
MMI of CLPOD.

\subsection{Equivalent Channel Model}
We want to calculate mutual information between input ${\bf u}$ of SSDD
or CLPOD encoder and output $R$ of MIMO channel, but in
Eq.~(\ref{eq:CHANNEL_MODEL1}), ${\bf u}$ is hidden in $S$. So we start with
extracting a relation between ${\bf u}$ and $R$ from
Eq.~(\ref{eq:CHANNEL_MODEL1}). The same derivation of the equivalent channel can
be found in~\cite{Hassibi}.

We define new vectors and new matrices as follows:
\begin{align*}
{\bf r}&=[{\bf r}_{R,1}^t,{\bf r}_{I,1}^t,\cdots,{\bf r}_{R,N}^t,{\bf r}_{I,N}^t]^t\in{\bf R}^{2TN},\\
{\bf w}&=[{\bf v}_{R,1}^t,{\bf v}_{I,1}^t,\cdots,{\bf v}_{R,N}^t,{\bf v}_{I,N}^t]^t\in{\bf R}^{2TN},\\
{\bf g}_n&=[{\bf h}_{R,n}^t,{\bf h}_{I,n}^t]^t,\\
B_q&=\left[
\begin{array}{@{\,}cc@{\,}}
A_{R,q} & -A_{I,q}\\
A_{I,q} & A_{R,q}\\
\end{array}
\right],\\
G&=
\left[
\begin{array}{@{\,}ccc@{\,}}
B_1{\bf g}_1 & \ldots & B_{2Q}{\bf g}_1\\
\vdots & \ddots & \vdots\\
B_1{\bf g}_N & \ldots & B_{2Q}{\bf g}_N\\
\end{array}
\right]\in{\bf R}^{2NT\times 2Q},
\end{align*}
where the vectors ${\bf x}_{R,n},\ {\bf x}_{I,n}$ denote the $n$th
column of real and imaginary part of the matrix $X$
respectively. Then Eq.~(\ref{eq:CHANNEL_MODEL1}) can be equivalently written
as
\begin{equation}
{\bf r}=\sqrt{\frac{\rho}{M}}G{\bf u}+{\bf w}. \label{eq:Channel_Real}
\end{equation}

\subsection{MMI of SSDD}
The equivalent channel matrix $G$ is known to the receiver because the
original channel matrix $H$ and the dispersion matrices $\{A_q\}$ are
known to receiver. Note also that the vector ${\bf r}$ is equivalent to
received block $R$. Therefore MMI of SSDD with $M$ transmit antennas,
$N$ receive antennas, $T$ block length, $Q$ complex information symbols
(i.e., $2Q$ real information symbols) and dispersion matrices $\{A_q\}$ at SNR $\rho$ is equal to
\begin{eqnarray}
\lefteqn{C_{\rm SSDD}(\rho,M,N,T,Q,\{A_q\})}&&\nonumber\\
&=&\frac{1}{T}\max_{p({\bf u}:\tr(D_A\Gamma_{\bf u})\leq TM)}I({\bf u};{\bf r},G), \label{eq:MMI_SSDD1}
\end{eqnarray}
where the factor $1/T$ normalizes the mutual information for the $T$
channel uses spanned by SSDD and $\tr(D_A\Gamma_{\bf u})\leq TM$ denotes
power constraint on ${\bf u}$. By the derivation similar to
~\cite{Telatar,Hassibi}, we can rewrite Eq.~(\ref{eq:MMI_SSDD1}) as
\begin{align}
&C_{\rm SSDD}(\rho,M,N,T,Q,\{A_q\})=\nonumber \\
&\frac{1}{2T}\max_{\Gamma_{\bf u}:\tr(D_A\Gamma_{\bf u})\leq TM}E_H\left[\log\det\left(I_{2Q}+\frac{2\rho}{M}G^tG\Gamma_{\bf u}\right)\right], \label{eq:MMI_SSDD2}
\end{align}
where the expectation is taken over the distribution of the original
channel matrix $H$. 
It is difficult to simplify Eq.~(\ref{eq:MMI_SSDD2}), but we can derive
its upper bound by recognizing that $\log\det(\cdot)$ is concave
function over the set of positive semi-definite matrices and using Jensen's inequality. We explain detailed derivation of upper bound as follows.

Since the covariance matrix $\Gamma_{\bf u}$ is positive semi-definite,
there is at least one square matrix $F=\sqrt{\Gamma_{\bf u}}$ that satisfies $FF^t=\Gamma_{\bf u}$.
So using determinant identity $\det(I_m+AB)=\det(I_n+BA),\ A\in {\bf R}^{m\times n},\ B\in {\bf R}^{n\times m}$, we rewrite Eq.~(\ref{eq:MMI_SSDD2}) as
\begin{eqnarray}
C_{\rm SSDD}&=&\frac{1}{2T}\max_{\Gamma_{\bf u}:\tr(D_A{\tilde
 \Gamma_{\bf u}})\leq
 TM}\nonumber\\
& &E_H\left[\log\det\left(I_{2Q}+\frac{2\rho}{M}F^tG^tGF\right)\right].\nonumber\\
& & \label{eq:MMI_SSDD9}
\end{eqnarray}
The term $\log\det(I_{2Q}+(2\rho/M)F^tG^tGF)$ in Eq.~(\ref{eq:MMI_SSDD9}) is a concave function of
$G^tG$ because for any positive semi-definite matrices $A,\ B\in R^{2Q\times 2Q}$ and
for any real number $0\leq \lambda\leq 1$, the inequality
\begin{eqnarray*}
\lefteqn{\log\det\left(I_{2Q}+\frac{2\rho}{M}F^t\left\{\lambda A+(1-\lambda)B\right\}F\right)}& & \nonumber \\
&\geq&\lambda\log\det\left(I_{2Q}+\frac{2\rho}{M}F^tAF\right)\nonumber\\
& &+(1-\lambda)\log\det\left(I_{2Q}+\frac{2\rho}{M}F^tBF\right) \label{eq:concave}
\end{eqnarray*}
holds. Therefore we may apply Jensen's inequality to
Eq.~(\ref{eq:MMI_SSDD9}) to obtain an upper bound of MMI of SSDD
\begin{eqnarray}
 C_{\rm SSDD}&\leq&\frac{1}{2T}\max_{\Gamma_{\bf u}:\tr(D_A{\tilde
 \Gamma_{\bf u}})\leq
 TM}\nonumber\\
& &\log\det\left(I_{2Q}+\frac{2\rho}{M}F^tE_H[G^tG]F\right).\nonumber\\
& & \label{eq:MMI_SSDD10}
\end{eqnarray}
The matrix $G^tG$ in Eq.~(\ref{eq:MMI_SSDD10}) is a function of the set of dispersion matrices $\{A_q\}$ as well as the channel matrix $H$. We can simplify $G^tG$ by using  a necessary and sufficient condition Eq.~(\ref{eq:SSDD}) for SSDD. From definitions, we have
\begin{equation}
G^tG=\sum_{n=1}^N\left[
\begin{array}{@{\,}ccc@{\,}}
{\bf g}_n^tB_1^tB_1{\bf g}_n & \ldots & {\bf g}_n^tB_1^tB_{2Q}{\bf g}_n\\
\vdots & \ddots & \vdots\\
{\bf g}_n^tB_{2Q}^tB_1{\bf g}_n & \ldots & {\bf g}_n^tB_{2Q}^tB_{2Q}{\bf g}_n\\
\end{array}
\right] \label{eq:MMI_SSDD11}
\end{equation}
and
\begin{eqnarray}
\lefteqn{{\bf g}_n^tB_q^tB_r{\bf g}_n}& &\nonumber\\
&=&{\bf h}_{R,n}^t\Re\left(A_q^HA_r\right){\bf h}_{R,n}+{\bf h}_{I,n}^t\Re\left(A_q^HA_r\right){\bf h}_{I,n}\nonumber\\
& &-{\bf h}_{R,n}^t\left\{\Im\left(A_q^HA_r\right)-\Im\left(A_q^HA_r\right)^t\right\}{\bf h}_{I,n}. \label{eq:QUADRATIC_FORM}
\end{eqnarray}
To simplify Eq.~(\ref{eq:QUADRATIC_FORM}), we use the following lemma.\\

\noindent
\textit{Lemma:}If the set $\{A_q\}$ of $2Q$ matrices satisfying Eq.~(\ref{eq:SSDD}) then following equalities hold for any real vectors ${\bf x},\ {\bf y}\in{\bf R}^M$:
\[
{\bf x}^t\Re(A_q^HA_r){\bf x}=0,\ 1\leq q\neq r\leq 2Q,
\]
\[
 {\bf x}^t\left\{\Im\left(A_q^HA_r\right)-\Im\left(A_q^HA_r\right)^t\right\}{\bf y}=0,\ 1\leq q,r\leq 2Q.
\]
\\

\noindent
\textit{Proof:}
\begin{eqnarray*}
{\bf x}^t\Re(A_q^HA_r){\bf x}&=&\frac{1}{2}\left\{{\bf
 x}^t\Re(A_q^HA_r){\bf x}+{\bf x}^t\Re(A_q^HA_r){\bf x}\right\}\\
&=&\frac{1}{2}{\bf x}^t\Re(A_q^HA_r+A_r^HA_q){\bf x}\\
&=&0
\end{eqnarray*}
\begin{eqnarray*}
\lefteqn{{\bf
 x}^t\left\{\Im\left(A_q^HA_r\right)-\Im\left(A_q^HA_r\right)^t\right\}{\bf y}}& &\\
&=&{\bf x}^t\left\{\Im\left(A_q^HA_r\right)+\Im\left(A_r^HA_q\right)\right\}{\bf y}\\
&=&{\bf x}^t\left\{\Im\left(A_q^HA_r+A_r^HA_q\right)\right\}{\bf y}\\
&=&0
\end{eqnarray*}
\hfill Q.E.D.\\
By using Lemma, we obtain
\begin{align}
G^tG=\sum_{n=1}^N\diag\left[\right.& {\bf h}_{R,n}^t\Re\left(A_1^HA_1\right){\bf h}_{R,n}\nonumber\\
&+{\bf h}_{I,n}^t\Re\left(A_1^HA_1\right){\bf h}_{I,n},\nonumber \\
&\cdots,\nonumber\\
&{\bf h}_{R,n}^t\Re\left(A_{2Q}^HA_{2Q}\right){\bf h}_{R,n}\nonumber\\
&\left.+{\bf h}_{I,n}^t\Re\left(A_{2Q}^HA_{2Q}\right){\bf h}_{I,n}\right]. \label{eq:MMI_SSDD12}
\end{align}
Denoting $m$th entries of ${\bf h}_{R,n}$ and ${\bf h}_{I,n}$ and $(l,m)$
entry of $\Re(A_q^tA_q)$ as $h_{R,n}(m),\ h_{I,n}(m)$ and $a_q(l,m)$,
respectively, we have
\begin{eqnarray}
\lefteqn{E_H\left[{\bf h}_{R,n}^t\Re(A_q^tA_q){\bf h}_{R,n}+{\bf h}_{I,n}\Re(A_q^rA_q){\bf h}_{I,n}\right]}& & \nonumber \\
&=&\sum_{m=1}^Ma_q(m,m)E\left[h_{R,n}(m)^2\right]\nonumber\\
& &+\sum_{l<m}^Ma_q(l,m)E\left[h_{R,n}(l)\right]E\left[h_{R,n}(m)\right] \nonumber \\
& &+\sum_{m=1}^Ma_q(m,m)E\left[h_{I,n}(m)^2\right]\nonumber\\
& &+\sum_{l<m}^Ma_q(l,m)E\left[h_{I,n}(l)\right]E\left[h_{I,n}(m)\right] \label{eq:DA1} \\
&=&\frac{1}{2}\sum_{m=1}^Ma_q(m,m)+\frac{1}{2}\sum_{m=1}^Ma_q(m,m) \label{eq:DA2} \\
&=&\tr\left[\Re(A_q^tA_q)\right] \nonumber \\
&=&\tr(A_q^HA_q),\nonumber
\end{eqnarray}
where Eq.~(\ref{eq:DA1}) follows from the statistical independence among channel
gains and Eq.~(\ref{eq:DA2}) follows from that the mean and variance of
channel gains $h_{R,n}(m),\ h_{I,n}(m)$ are 0 and 1/2, respectively. Therefore we have
\begin{equation}
 E_H\left[G(H)^tG(H)\right]=D_A. \label{eq:GtG}
\end{equation}
Substituting Eq.~(\ref{eq:GtG}) into Eq.~(\ref{eq:MMI_SSDD10}), we obtain
\begin{equation}
C_{\rm SSDD}\leq\frac{1}{2T}\max_{\Gamma_{\bf u}:\tr(D_A{\tilde
 \Gamma_{\bf u}})\leq
 TM}\log\det\left(I_{2Q}+\frac{2\rho N}{M}D_A{\tilde\Gamma}_{\bf
 u}\right). \label{eq:MMI_SSDD13}
\end{equation}
We simply choose $(TM/2Q)I_{2Q}$ as $D_A\Gamma_{\bf u}$ to
maximize the term $\log\det(\cdot)$ in Eq.~(\ref{eq:MMI_SSDD13}) and get
\begin{equation}
C_{\rm SSDD}\leq\frac{Q}{T}\log\left(1+\rho N\cdot\frac{T}{Q}\right). \label{eq:MMI_SSDD14}
\end{equation}
The maximum symbol rate $Q/T$ of SSDD for $M$ is unknown unlike CLPOD~\cite{Wang} or COD~\cite{Wang,Liang}. In section 4,
we numerically evaluate Eq.~(\ref{eq:MMI_SSDD14}) and show necessary
rate of SSDD at which MMI of SSDD can utilize full channel capacity. In
Section~\ref{Sec.4}, we search the tightness of the upper
bound~Eq.(\ref{eq:MMI_SSDD14}) by comparing it with delay optimal COD at the same symbol
rate.
\subsection{An expression of MMI of CLPOD}
Unlike SSDD, we can derive an exact expression of
MMI of CLPOD due to the additional condition
Eq.~(\ref{eq:CLPOD}). Sandhu and Paulraj~\cite{Sandhu} derived an expression MMI of
CLPOD. We give alternative derivation of MMI of CLPOD
by using the result of MMI of SSDD in previous subsection. We also show that
MMI of CLPOD relates the capacity of another channel whose parameter
setting (i.e., number of transmit/receive antennas and SNR at each
receive antenna) differs from original one. In~\cite{Sandhu}, MMI for
given channel realization $H$ is computed but we compute average MMI
over $H$.

Denoting MMI of CLPOD as $C_{\rm CLPOD}$, we have
\begin{eqnarray}
\lefteqn{C_{\rm CLPOD}}& &\nonumber\\
&=&\frac{1}{2T}\max_{\Gamma_{\bf u}:\tr(\Gamma_{\bf u})\leq M}\nonumber\\
& &E_H\left[\log\det\left(I_{2Q}+\frac{2\rho}{M}G^tG\Gamma_{\bf u}\right)\right] \label{eq:MMI_CLPOD1} \\
&=&\frac{1}{2T}\max_{\Gamma_{\bf u}:\tr(\Gamma_{\bf u})\leq M}\nonumber\\
& &E_H\left[\log\det\left(I_{2Q}+\frac{2\rho}{M}\sum_{m=1}^M\sum_{n=1}^N|h_{mn}^2|\Gamma_{\bf u}\right)\right] \nonumber\\
& &\label{eq:MMI_CLPOD2} \\
&=&\frac{1}{2T}E_H\Bigg[\log\det\Bigg(I_{2Q}\nonumber\\
& &+\frac{2\rho}{M}\sum_{m=1}^M\sum_{n=1}^N|h_{mn}^2|\cdot\frac{M}{2Q}I_{2Q}\Bigg)\Bigg]
 \nonumber \\
&=&\frac{Q}{T}E_H\left[\log\left(1+\frac{\rho}{Q}\sum_{m=1}^M\sum_{n=1}^N|h_{mn}|^2\right)\right] \nonumber
\end{eqnarray}
where the first equality Eq.~(\ref{eq:MMI_CLPOD1}) follows from
Eq.~(\ref{eq:MMI_SSDD2}) and the second equality Eq.~(\ref{eq:MMI_CLPOD2}) follows from using additional condition Eq.~(\ref{eq:CLPOD}) of CLPOD in Eq.~(\ref{eq:MMI_SSDD12}). Note that MMI of CLPOD does not depend on the choice of the set of dispersion matrices $\{A_q\}$.
On the other hand, the capacity of MIMO fading channel with $MN$
transmit antennas and 1 receive antenna at SNR $MN\rho/Q$ equals to
\begin{equation}
E\left[\log\left(1+\frac{\rho}{Q}\sum_{i=1}^{MN}|h_i|^2\right)\right]\label{eq:CHANNEL_CAPACITY},
\end{equation}
where $h_i\ (i=1,\cdots,MN)$ are statistically independent complex
Gaussian random variable with zero mean and unit
variance~\cite{Telatar,Foschini}. Therefore denoting channel capacity
with $M'$ transmit antennas and $N'$ receive antennas at SNR $\rho'$ as $C(\rho',M',N')$, we have a relation of MMI of
CLPOD and channel capacity
\begin{equation}
C_{\rm CLPOD}(\rho,M,N,T,Q)=\frac{Q}{T}C\left(\frac{MN}{Q}\rho,MN,1\right)\label{eq:MMI_CLPOD3}.
\end{equation}
From Eq.~(\ref{eq:MMI_CLPOD3}), we have three important observations.
\begin{itemize}
\item It is well known that, at high SNR, the channel capacity is mainly
      dominated by the value of
      $\min\{M,N\}\log\rho$~\cite{Telatar,Foschini}. On the other hand,
      the third argument of the right hand of Eq.~(\ref{eq:MMI_CLPOD3})
      is 1, which corresponds to the number of receive
      antennas. Therefore increasing numbers of both transmit and
      receive antennas does not increase MMI of CLPOD.
\item $C_{\rm CLPOD}$ is proportional to symbol rate $Q/T$. However it
      is known that for more than three transmit antennas, maximum
      symbol rate achieved by CLPOD is equal to or less than 3/4~\cite{Wang}.
\item Under the condition that the symbol rate is constant, the delay optimal code i.e., the code
      having the minimum block length $T$ and the minimum number $Q$ of complex symbols has largest
      MMI.
\end{itemize}

\section{Numerical Result}
\label{Sec.4}
\begin{figure}[t]
\begin{center}
\includegraphics*[width=\linewidth]{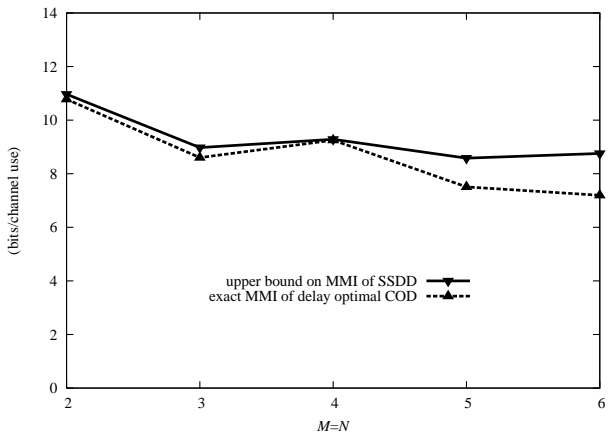}
\caption{Comparison between the upper bound of MMI of SSDD and exact MMI
 of delay optimal COD for 2 to 6 antennas at $\rho=30$[dB]}
\label{figure1}
\end{center}
\end{figure}

To show the tightness of Eq.~(\ref{eq:MMI_SSDD14}), the upper
bound of MMI of SSDD and exact MMI of delay optimal COD are shown in
Fig.~\ref{figure1}. We set SNR $\rho$ at 30[dB] and use
parameters in Table~\ref{table1} for computation, where design parameters $Q,\ T$ for COD is delay
optimal one~\cite{Liang} and the symbol rate of SSDD is same as that of
COD. Fig.~\ref{figure1} shows that under the same rate condition, the
upper bound of MMI of SSDD is close to the exact MMI of COD for 2 to 4
transmit/receive antennas especially. Therefore we may regard
Eq.~(\ref{eq:MMI_SSDD14}) as a tight upper bound on MMI of SSDD.

We give the necessary symbol rate of SSDD at
which MMI of SSDD can achieve channel capacity at $\rho=$10,\ 20 and 30[dB] for
2 to 8 transmit/receive antennas, in Figure 2. We see that this value
linearly increases as $M$ and this value is needed to be much larger
than that of CLPOD upper bounded by 3/4.

\begin{table}[t]
\begin{center}
\begin{tabular}{|c|c|c|c|c|} \cline{2-5}
\multicolumn{1}{c|}{} & \multicolumn{1}{c|}{SSDD} & \multicolumn{3}{c|}{COD} \\\hline
$M=N$ & $Q/T$ & $Q$ & $T$ & $Q/T$ \\\hline
2 & 1 & 2 & 2 & 1 \\\hline
3 & 3/4 & 3 & 4 & 3/4   \\\hline
4 & 3/4 & 3 & 4 & 3/4   \\\hline
5 & 2/3 & 10 & 15 & 2/3 \\\hline
6 & 2/3 & 20 & 30 & 2/3 \\\hline
\end{tabular}
\end{center}
\caption{Parameter setting of SSDD and COD for 2 to 6 antennas}
\label{table1}
\end{table}

\begin{figure}[t]
\begin{center}
\includegraphics*[width=\linewidth]{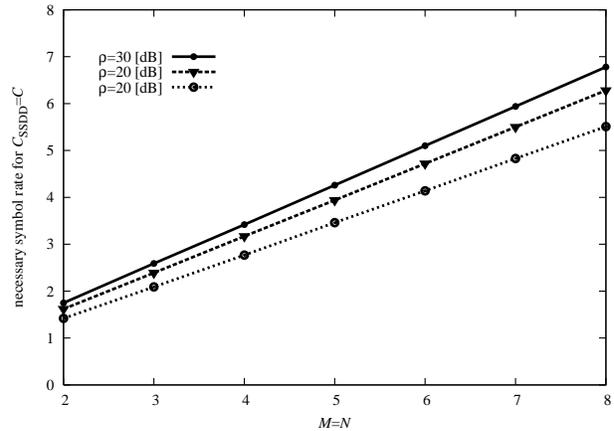}
\caption{Necessary symbol rate for $C_{\rm SSDD}=C$ for 2 to 8
 transmit/receive antennas}
\label{figure2}
\end{center}
\end{figure}

\section{Conclusion}
\label{Sec.5}
In this paper we gave a tight upper bound of MMI of SSDD and alternative
derivation of an exact
expression of MMI of CLPOD. We showed the necessary symbol rate of SSDD at
which MMI of SSDD can attain channel capacity and these value are much
larger than the symbol rate of CLPOD. To find out more about performance
of SSDD, the research of the maximum symbol rate of SSDD is needed.

\end{document}